\documentclass[letter,scriptaddress,twocolumn, showkeys, article]{revtex4}
\usepackage{ amssymb }
	\usepackage{amsmath}
	\usepackage{makeidx}
	\usepackage{subfigure}
	\usepackage{amsfonts}
	\usepackage[ansinew]{inputenc}
	\usepackage[usenames,dvipsnames]{pstricks}
	\usepackage{subfigure}
	\usepackage{epsfig}
	\usepackage{pst-grad} 
	\usepackage{pst-plot} 
		\usepackage{amsthm}
\usepackage{ amssymb }
 	\usepackage{makeidx}
	\usepackage{amsfonts}
 	\usepackage{listings}
 	\usepackage{mathtools}
	\usepackage{float}
	\usepackage{amsmath,amssymb,amsfonts}
	\usepackage[colorlinks,hyperindex]{hyperref}
	\hypersetup
	{
		colorlinks,%
		citecolor=blue,%
		linkcolor=blue,%
		urlcolor=blue,%
	}
	\usepackage{tikz}

\usepackage{amsmath}
\usepackage{appendix}
\usepackage{amssymb}
\usepackage{graphicx}
\usepackage{xcolor}
\usepackage{float,graphicx}
\usepackage{placeins}
\usepackage{color, colortbl}
\usepackage[colorinlistoftodos]{todonotes}
\graphicspath{ {./Images and Graphs/} }

\theoremstyle{definition}

	\theoremstyle{plain}

\makeindex

\begin{document}

\title{  Correlations Between COVID-19 and Dengue}

\author{ Paula Bergero$^{a}$,  Laura P. Schaposnik$^{\star, b}$, Grace Wang$^{c}$}
  \affiliation{($\star$) Corresponding author: schapos@uic.edu}

\begin{abstract}
A dramatic increase in the number of outbreaks of Dengue has  recently been reported, and climate change is likely to extend the geographical spread of the disease. In this context, this paper shows how a neural network approach can incorporate Dengue and COVID-19 data as well as external factors (such as social behaviour or climate variables), to develop predictive models that could improve our knowledge and provide useful tools for health policy makers. Through the use of neural networks with different social and natural parameters, in this paper we define a {\it Correlation Model} through which we show that the number of cases of COVID-19 and Dengue have very similar trends. We then illustrate the relevance of our model by extending it to a Long short-term memory model (LSTM)  that incorporates both diseases, and using this to  estimate Dengue infections  via COVID-19 data   in countries that lack sufficient Dengue data. 
 \end{abstract}

\maketitle
 
 \section{Introduction} \label{Introduction}
 Since its emergence, COVID-19 has been studied in correlation with many other infectious diseases, most notably influenza, meningitis, and measles \cite{Berberian}. Caused by a coronavirus known as SARS-CoV-2, COVID-19 has become a worldwide pandemic that has, as of April of 2022,  over 500 million confirmed cases and over six million deaths have been reported globally by  \href{https://www.who.int/publications/m/item/weekly-epidemiological-update-on-covid-19---27-april-2022}{WHO}.
 Although a large amount of research has been done both on Dengue as well as on COVID-19, not much work has been done on their correlation, and thus this shall be the focus of the present paper. In particular, building on \cite{Bergero}, we study such correlation worldwide using machine learning techniques, and explore different directions in which the results we deduce could be used by policy makers in the future, especially since the syndemic and possible simultaneous transmission presents a significant challenge to public health efforts in those areas \cite{corre}.

Prominent in Latin American and Asian countries, Dengue is a viral disease that is transmitted by mosquitoes, especially those of species Aedes aegypti (Aa) \cite{paho}. Dengue is a highly seasonal, multi-annual disease, most prevalent before and after rainy seasons. 
Individuals who recover from Dengue have long-term immunity against that specific serotype (homotypic), and they also have short-term immunity against Dengue of a different serotype (heterotypic) \cite{Dengue_Immunity}. Moreover, Dengue is widely considered to be the most important mosquito-borne viral disease and is very wide spread, covering mostly tropical and sub-tropical areas, between the January isotherm and the July isotherm of 10 degrees Celsius. Temperature, rainfall, and population density were all factors shown to be associated with the number of Dengue infections, which explains the regions where it is most prevalent \cite{global_Dengue}.

Recently, infections of Dengue and COVID-19 have begun to be considered together, particularly in South America. Despite the year 2020 being an epidemiological complex year with the pandemic of COVID-19, the fatality rate from Dengue was recorded as 0.04\%, the lowest in the past decade \cite{paho}. Interestingly, however, there seems to be a persistence of higher than expected number of Dengue cases in endemic areas, occurring simultaneously with intense COVID transmission.
Although it has been brought up in September of 2021 that there may be cross-immunity between the two diseases \cite{10.1093/cid/ciaa1895}, a population-based cohort study in December of 2021 found that individuals with prior dengue actually had an increased risk of clinically apparent COVID-19 \cite{increasedRisk}. Thus, there is likely not any cross-immunity between the two diseases.

We shall organize our paper in the following way: after introducing some background  in Section \ref{Background}, we define different external variables that we then use in a neural network model for Brazil in Section \ref{Impact of Parameters on COVID-19 in Brazil}. In particular, we consider holidays in Subsection \ref{Holiday Impact on COVID-19 Cases in Brazil} and  climate factors such as temperature and humidity in Subsection \ref{Impact of Humidity and Temperature in Brazil}. By adding in these factors, we 
\begin{itemize}
\item are able to {\bf create a model that predicts and correlates COVID-19 to}
\begin{itemize}
\item {\bf holidays};
\item {\bf climate,} and 
\item {\bf Dengue}.
\end{itemize}
\end{itemize}

After introducing and describing our novel model, we show its utility by applying it to  other nations or regions in Section \ref{Repeating the Model}. In particular, we use our model to study the data from  Peru and Colombia and 
\begin{itemize}
\item show the {\bf model's predictive qualities} by using COVID-19 data  to predict Dengue in countries that lack sufficient Dengue data. 
\end{itemize}
Other factors such as Latent Heat Flux and   peaks correlations are also considered in Section \ref{Additional Ideas}. 
Finally, since our Machine Learning model focused solely on external variables, we take a more time-series based approach and 
\begin{itemize}
\item introduce {\bf a  Long short-term memory  model (LSTM)} in Section \ref{LSTM}, obtaining a more comprehensive look at predicting future values for both diseases than does the standard neural network. 
\end{itemize}
The results and observations drawn from our work are summarized in Section \ref{conclusion}, where future directions of research are also outlined.
\pagebreak
 
 \section{Background} \label{Background}\label{ML}
 
 Machine Learning has been incredibly useful in countless topics, and COVID-19 is no exception. Ever since the beginning of the pandemic, researchers have been using machine learning to predict future outbreaks or possible correlations with other factors. In the present paper  we are specifically looking at the correlation of COVID-19 with Dengue. Hence, we shall dedicate this section to introducing the background needed in Machine Learning, neural networks and  LSTM models.

 \subsection{ Machine Learning and  COVID-19} \label{use}

Because Machine Learning is such a powerful tool in making predictions and finding correlations, researchers have been using it in studies about COVID-19 since the beginning of the pandemic. As early as April of 2020, in fact, researchers were finding that specific models, such as MLP (Multi-layered perceptron) and ANFIS (adaptive network-based fuzzy interference system) -- both artificial neural networks -- had promising results for estimating the severity of the pandemic. In the beginning of 2020, researchers also tested several models to estimate time-series data, finding that whilst the logistic model outperformed the others,  these fitted models were hard to generalize after  a 30-day period \cite{Ardabili2020.04.17.20070094}. More recently,  Machine Learning begun to be used to classify different types of COVID-19 since   it can also test different aspects, such as the severity of someone's COVID-19 symptoms. In particular, researchers have suggested a hybrid machine learning/deep learning model that predicts COVID-19 severity from CT images \cite{CT}.

\begin{figure}[h!]
    \centering
    {\includegraphics[width=0.35\textwidth]{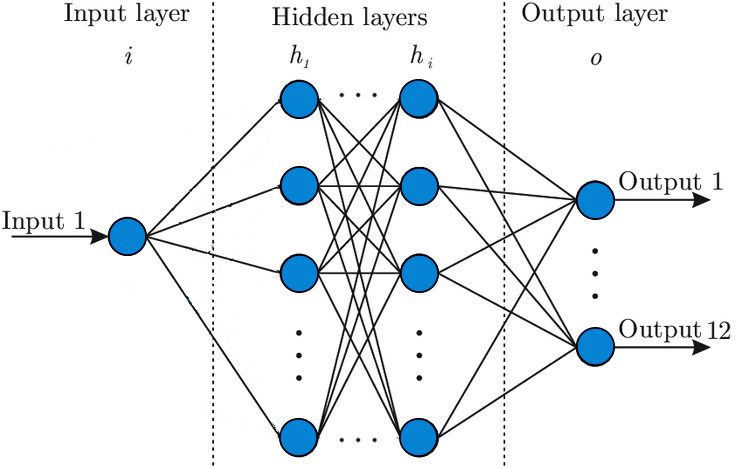}}
    \caption{An example of a neural network with $i$ hidden layers, 1 input layer and 12 output layers. }
    \label{network}
\end{figure}

Our work on the correlation between Dengue and COVID-19 will be done firstly by using a neural network to study data that was collected of COVID-19 and Dengue. By  incorporating parameters that are possibly related to these two infectious diseases  such as holiday seasons, temperature, or rainfall, we are able to draw strong correlations that then serve to build a predictive model. These models then shall generate loss curves and prediction graphs to find out what factors are the most prominent in these disease cases. 
A standard neural network has   input layers,   output layers, and one or more hidden layers as shown in \autoref{network}. 

  The learning structure used   here   has one input layer, four output layers using the relu activation function, and an output layer using the linear activation function.   The predicted infection number is given by the  output node with the greatest activation. The neural network was trained using
a training set of size 10 epochs and testing using a cross-
validations set of size approximately 150 epochs. To understand the model we consider its loss function using the Mean Absolute Error function in Section \ref{Impact of Parameters on COVID-19 in Brazil}, which was calculated by   averaging the absolute distances between the predicted and actual values. 
Data for Dengue fever has been collected from the \href{https://www3.paho.org/data/index.php/en/mnu-topics/indicadores-dengue-en/dengue-nacional-en/252-dengue-pais-ano-en.html}{Pan-American Health Organization Dataset on Dengue}, and data for COVID-19 has been collected from the World Health Organization \href{https://covid19.who.int/info}{dashboard on COVID-19}.

\subsection{LSTM Models} \label{LSTM_intro}

The correlation between Dengue and COVID-19 will also be studied through a  Recurrent Neural Network (RNN), in an effort to predict future values of the number of infections based on past observed data. Specifically, we shall consider a LSTM model  (Long short-term memory), which is particularly  useful especially in cases where there may be time lags, and which is often used in analyzing stock prices and the like. In the setting of our paper, we use the LSTM model to generate a model for each disease based solely on its past observed values and then we incorporate the time series values for the other disease as well as other external factors.

\begin{figure}[h!]
    \centering
    {\includegraphics[width=0.35\textwidth]{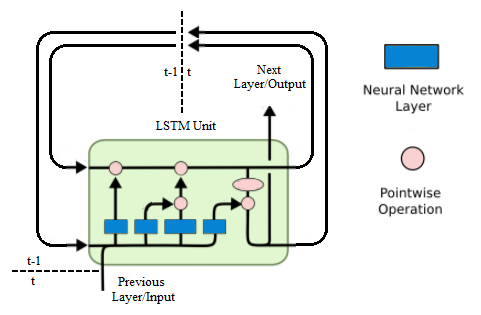}}
    \caption{An example of an LSTM model.}
    \label{network2}
\end{figure}

The LSTM model is more complicated than a standard neural network in that it can process sequences of data and not solely datapoints. Because these recurrent types of neural networks have ``loops'' (hence, recurrent) around them, the information is able to persist, something that standard neural networks cannot necessarily do. Essentially, a LSTM model contains a cell, input gate, and output gate as shown in   \autoref{network2}. But it also consists of a forget gate, something that traditional neural networks do not have. The input gate inserts new information into the cell, and the output gate passes along the new, updated information. The forget gate, on the other hand, forgets useless information, which is especially important because LSTM is used for long-term processing.
In our LSTM model, we have an input layer, a single hidden layer, and an output layer.

  \onecolumngrid

\begin{figure*}
    \centering
 \includegraphics[width=\textwidth]{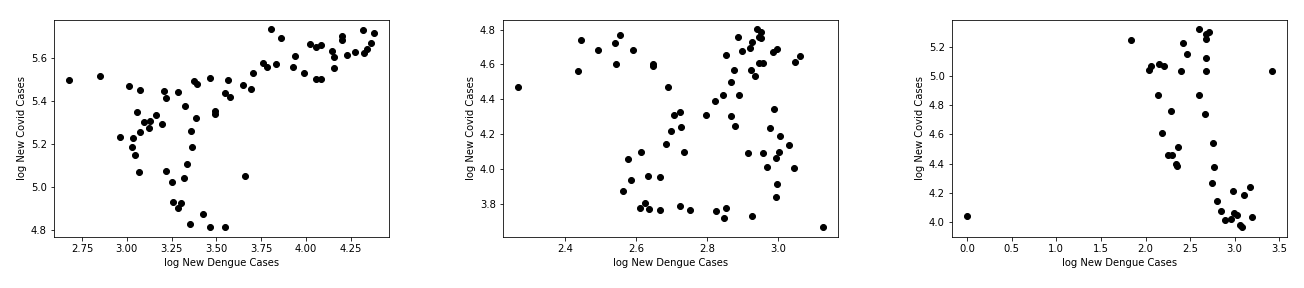}
    \caption{Dengue and COVID-19 cases plotted for Brazil  (left), Peru (middle), Colombia  (right).}
    \label{fig:correlations}
\end{figure*}
 \twocolumngrid

\section{Correlations via datasets}

In order to understand the data sets from the \href{https://www3.paho.org/data/index.php/en/mnu-topics/indicadores-dengue-en/dengue-nacional-en/252-dengue-pais-ano-en.html}{Pan-American Health Organization Dataset on Dengue} and the World Health Organization \href{https://covid19.who.int/info}{dashboard on COVID-19} which we will consider in this paper, we shall build some correlation plots to gauge the data at hand. 

We shall begin our work with the largest country in South America, and the most prominent with respect to COVID-19 and Dengue, which is Brazil. Since the number of COVID-19 cases is relatively large, one may take the base 10 logarithms of all the data to understand the sets better: a graph of the base 10 logarithm for new COVID-19 cases per week vs the base 10 logarithm for new Dengue cases per week can be seen in   \autoref{fig:correlations}  (Left) above.

Along this paper we shall  only used data from epidemiological weeks 30 to 100, which corresponds to the count beginning in January  2020. This is done since outside of that range  the datasets are  too irregular, and thus including those values would not be representative of the actual trend. Through  \autoref{fig:correlations} one can deduce some initial properties of the correlation between Dengue and COVID-19 in South America.   In the correlation graph for Brazil in  \autoref{fig:correlations} (Left), one can see that the two diseases have a roughly positive correlation. A similar study of the datasets for Peru and Colombia are presented in \autoref{fig:correlations} (Middle) and (Right) respectively.

\begin{figure}[H]
\begin{center}
  \includegraphics[scale = 0.55]{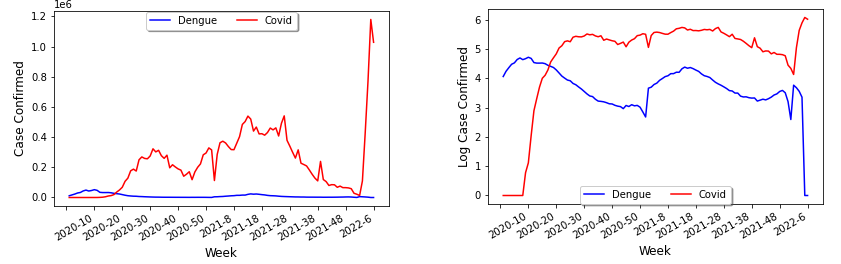}
  \end{center}
  \caption{Plot of data from the \href{https://www3.paho.org/data/index.php/en/mnu-topics/indicadores-dengue-en/dengue-nacional-en/252-dengue-pais-ano-en.html}{Pan-American Health Organization Dataset on Dengue} and the World Health Organization \href{https://covid19.who.int/info}{dashboard on COVID-19} (left) and its log plot (right).}
  \label{fig:Brazil-threeYears}
\end{figure}

It is interesting to note  that the graph for Peru is very random, whereas the graph for Colombia shows slightly negative correlation, suggesting that the notion that the correlation between the diseases could differ by country.  
Returning the attention to Brazil, we shall consider a graph from the years 2020-2022 for Dengue and COVID-19 infections in Brazil, where the horizontal axis is the epidemiological week, counted from the beginning of 2020, and the  vertical axis represents the number of cases in millions as shown in \autoref{fig:Brazil-threeYears}. One can then compare the number of  COVID-19 cases (red) and   the number of Dengue cases (blue), as shown in  \autoref{fig:Brazil-threeYears} (Left), noting that  between week 60 and week 80, the peak of COVID-19 generally coincides with the peak of Dengue, a period that corresponds to the year 2021. 
After week 100 the number of COVID-19 cases spikes up, likely due to the recent increases of Omicron. In \autoref{fig:Brazil-threeYears} (Left) the data has not been normalized, and thus one can see   the number of COVID-19 cases being much greater than the number of Dengue cases.  

Since the number of COVID-19 cases is of a greater magnitude than the number of Dengue cases, it becomes difficult to see a definite trend. It is thus useful to consider the base 10 log of the number of COVID-19 and Dengue cases as shown in \autoref{fig:Brazil-threeYears} (Right) to note that the peaks of Dengue generally coincide with the peaks of COVID-19. 

 \onecolumngrid

 \twocolumngrid

Through \autoref{fig:Brazil-threeYears} (Right) in the previous page, one can see that from approximately from week 30 to week 100 the the increase or decrease of cases for COVID-19 and Dengue are correlated. In particular, one  can see that in the beginning of year 2020, there was a much greater number of Dengue infections than COVID-19 cases, since the pandemic had not emerged yet. After week 100, or roughly the beginning of year 2022, the Omicron variant resulted in the number of COVID-19 cases increasing dramatically. 
Interestingly,  both the number of COVID-19 and Dengue   experience a drop around weeks 50 and 100 (during 2021). Whilst this may be due to the fact that the period coincides with the wintertime and  the end of the year, this may be counterintuitive because one would think that infectious diseases  would spike during holiday times.

\begin{figure}[H]
\begin{center}
  \includegraphics[scale = 0.6]{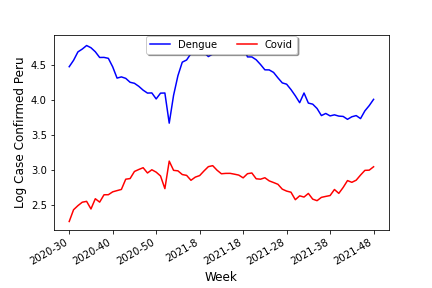}
  \end{center}
  \caption{Weekly correlations between COVID-19 and Dengue for Per\'u.}
  \label{fig:Peru_correlation}
\end{figure}

A similar study to the one done above can be performed for the dataset of Per\'u, leading to \autoref{fig:Peru_correlation} which shows  the data against the epidemiological weeks for Per\'u, where   the number of cases for each disease  seem to follow the same general trend. 

In order to understand other parameters which might be influencing the cases of both COVID-19 and Dengue, we shall consider the particular influence of humidity and temperature.  Following the style from the previous analysis, we can draw the explicit correlations between the various parameters and the number of cases of either COVID-19 or Dengue. In particular \autoref{fig:Peru_parameterCorrelations}  below shows the correlation between COVID-19/Dengue and Temperature/Humidity, where the black dots represent the points for COVID-19, and the red dots represent the points for Dengue. By analyzing the graphs one can deduce that temperature would be a better predictor than humidity for the number of cases.

\bigskip

\section{A Correlation Model} \label{Impact of Parameters on COVID-19 in Brazil}

In order to study the correlation between COVID-19 and Dengue in further detail, we shall build a neural network, {\it  the Correlation Model}, for our model (see  Section \ref{Background} for background). Our model has an input layer consisting of the various parameters we add, four hidden layers within the network, and an output layer.

As with the previous section, we shall begin our study with the dataset from Brazil. The Correlation Model will generates a prediction for the number of COVID-19 cases, once additional   variables are taken into consideration. We have built the first version of the model to include:   \begin{itemize}
\item the number of Dengue infections, 
\item the boolean variable of whether the week contains a ``holiday'', and 
\item quantitative climate factors (
  temperature,   humidity, and 
  rainfall). 
 
\end{itemize}

In what follows we shall describe the role each variable plays within the correlation study, and how we are able to incorporate such variables into our Correlation Model.

 \onecolumngrid

\begin{figure}[H]
\begin{center}
  \includegraphics[scale = 0.59]{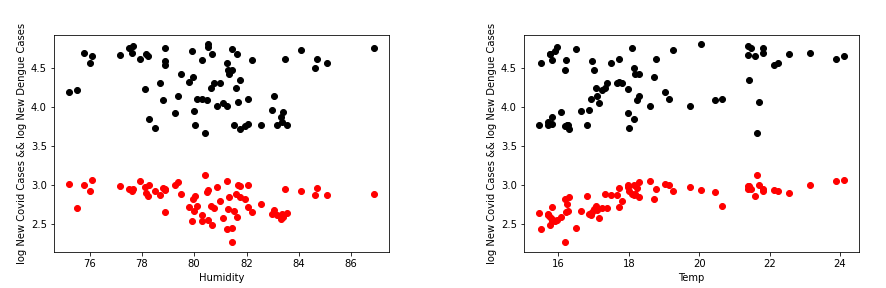}
  \end{center}
  \caption{Humidity Correlations (Left) and Temperature Correlations (Right) for Per\'u's dataset. }
  \label{fig:Peru_parameterCorrelations}
\end{figure}

 \twocolumngrid
 \subsection{Impact of Holidays} \label{Holiday Impact on COVID-19 Cases in Brazil}

\begin{figure}[H]
  \includegraphics[scale = 0.62]{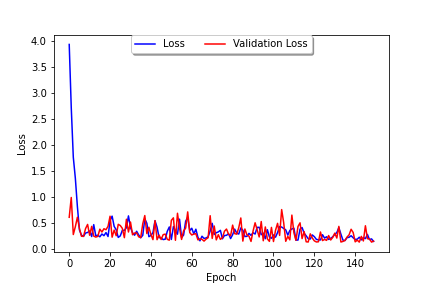}
  \caption{A graph of the loss curve when the holiday data is first added.}
  \label{fig:holiday_loss}
\end{figure}

The impact of public holidays on epidemic spreads has long been studied, from generic epidemic standpoints \cite{holiday2} to directly related to the latest COVID-19 epidemic \cite{holiday1}. It is thus natural to consider how holidays influence the correlation between COVID-19 and Dengue, and to study this question we shall consider the Brazilian dataset and a parameter which tracks whether or not a week contains an important Brazilian holiday. The list of holidays used can be found in Appendix \ref{appendix_1}, and our parameter has a value of 1 during the  weeks that contain the holidays in 2020, 2021, and 2022, and $0$ otherwise. 
 We firstly study the loss curve for this model, plotted in \autoref{fig:holiday_loss}, where    the blue line represents the training loss and the red line represents the validation loss. As in usual neural networks, the data has been split randomly into a training set and a cross-validation set, with our epoch size of 150 and the loss function   based on the Mean Absolute Error.  

\begin{figure}[H]
\begin{center}
  \includegraphics[scale = 0.6]{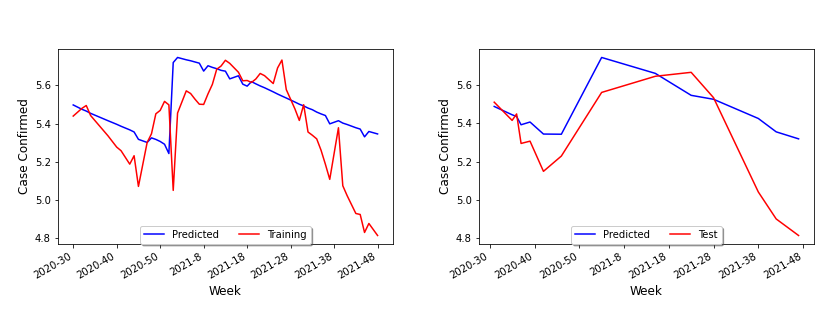}
  \end{center}
  \caption{A graph of the actual data and the predicted data  for holidays on the training set (Left) and the test set (Right).}
  \label{fig:holiday_test}
\end{figure}

Since one of the most useful things that the Keras model can do is predict new values, we used model.predict() to do so within our setting. Specifically, our Correlation Model then predicts the number of cases for COVID-19 leading to the visualization in  \autoref{fig:holiday_test} (Left),   where the blue line is the predicted value for any given week using the prediction algorithm, and the red line is the actual value in the dataset.

Notice that the vertical axis runs from 4.8 to 5.6. Recall that we have taken the base-10 logarithm of the weekly cases for COVID-19 and Dengue as a means of normalization and for the number of cases to be of the same approximate magnitude. In   \autoref{fig:holiday_test} (Left),  the prediction has the same general trend as the actual data and can therefore indicate where the approximate peaks and valleys will be. Yet, the prediction has a smaller amplitude than the actual data, and thus  the prediction would seem to  diminish the variance of the initial data.  To account for this,  we can define a new variable $C$ to be the \emph{Contraction ratio}: the ratio of the average distance from the mean for the actual data to the average distance from the mean for the predictions. We shall se the importance of this variable in the upcoming sections as well as in the conclusion. 

In order to test our model, one   repeats the process for the prediction done before and generates a new test graph shown in  \autoref{fig:holiday_test} (Right), which presents the study on the cross-validation set, not the training set. In this case, the blue line represents the prediction model on the validation set, and the red line is the actual data in the validation set.
Since not as many weeks were included in the test set in   \autoref{fig:holiday_test} (Right), the relationship between the two curves is less apparent than that of   \autoref{fig:holiday_test} (Left). However, one can still see that the prediction and the actual data for the validation set have the same approximate trend, and that like in the training set, the prediction values do not vary as much as the actual data values.  \smallbreak
 
 \onecolumngrid

 \twocolumngrid

  \bigskip

\subsection{Impact of Humidity and Temperature} \label{Impact of Humidity and Temperature in Brazil}
Research has  shown that climate \cite{climate1} and in particular increasing temperatures influence COVID-19 transmission \cite{NOTARI2021144390}, as well as for other viruses \cite{climate2}. Therefore, in what fallows we shall add climate factors such as temperature or humidity to our model. A characteristic of Dengue is that it is most prevalent around rainy seasons, and thus one should expect higher value of humidity would correspond to a greater number of Dengue cases and we shall look into this correlation below. 

\begin{figure}[H]
\begin{center}
  \includegraphics[scale = 0.45]{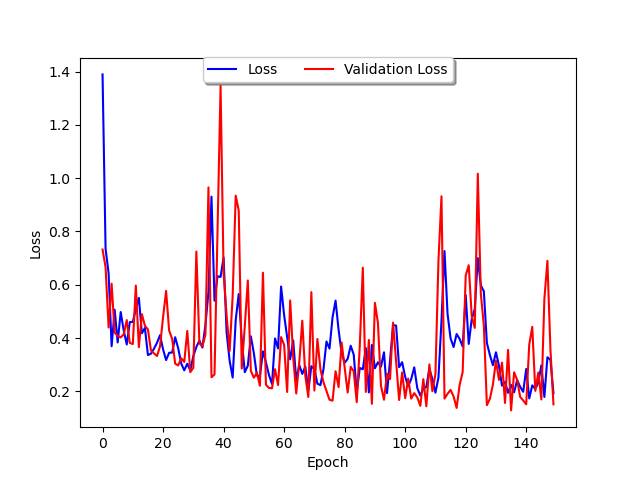}
  \end{center}
  \caption{The loss curve for only climate factors.}
  \label{fig:climateOnly_loss}
\end{figure}

 The dataset obtained on Brazilian climate contains daily maximum and minimum temperature and humidity for major Brazilian cities, so in order to fit it to our model, we can take the average of the temperature and humidity over each epidemiological week. In addition to the factors of temperature and humidity, we shall weight the data based on population, so that the greater the population, the more heavily it is weighted. To perform our study we consider the two  biggest cities in Northern and Southern Brazil, which are  Salvador and Sao Paulo, respectively. The weighted   data according to the populations of Northern and Southern Brazil, is given by approximately 43\% and 57\%, respectively, leading to the loss curve model shown in \autoref{fig:climateOnly_loss}.

\begin{figure}[H]
\begin{center}
  \includegraphics[scale = 0.45]{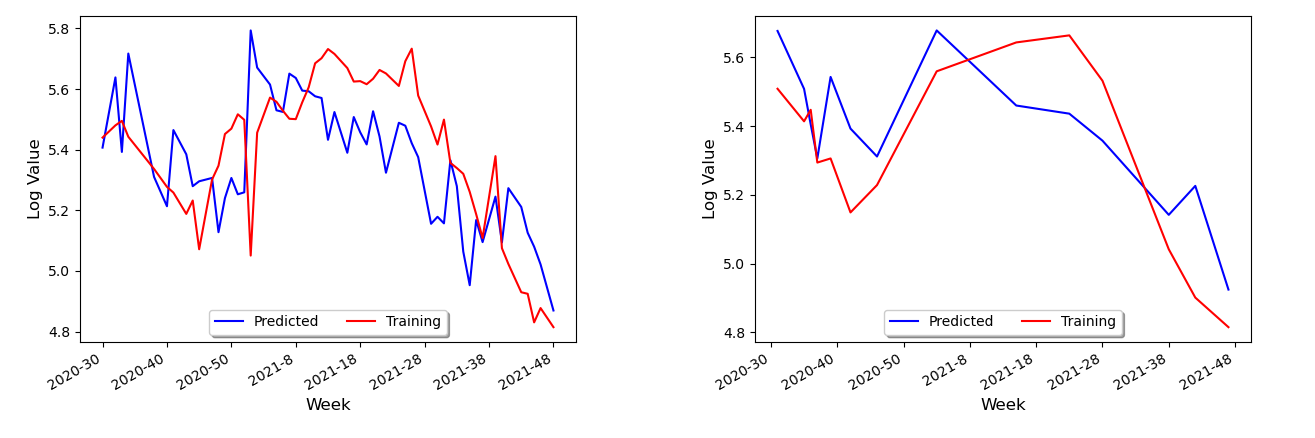}
  \end{center}
  \caption{Actual data and   predicted data for only climate on the training set (Left) and the test set (Right).}
  \label{fig:climateOnly_test}
\end{figure}

Unlike the study done in Figure \ref{fig:holiday_loss}, one can observe that the loss is generally greater, and thus the model generates less precise results, leading to the conclusion that  that climate factors are not as correlated with COVID-19 cases as holiday factors are.
Moreover,  a graph of the predicted data based on our neural network model on the training set can be obtained, as shown in \autoref{fig:climateOnly_test} (Left) below. In particular, one can see that the model  predicts the actual numbers for COVID-19 rather well, as the two graphs have approximately the same amplitude and the same mean. Finally, one can repeat this for the test set, and find that, similar to before, the model in \autoref{fig:climateOnly_test} (Right)  predicts the general trend of ups and downs fairly well for the test set.

Having studied our Correlation Model with climate parameters and holiday parameters separately, we shall now consider both sets of parameters together, leading to the comparison chart featured in \autoref{fig:climate_loss} which describes the loss curve for both climate and holidays:

\begin{figure}[H]
  \includegraphics[scale = 0.55]{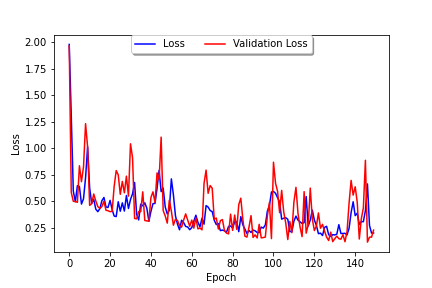}
  \caption{The loss curve for both climate and holidays. Data on the number of reported cases is from \href{https://www3.paho.org/data/index.php/en/mnu-topics/indicadores-dengue-en/dengue-nacional-en/252-dengue-pais-ano-en.html}{Pan-American Health Organization Dataset on Dengue} and the World Health Organization \href{https://covid19.who.int/info}{dashboard on COVID-19}. Data on Brazilian climate is from \href{https://www.visualcrossing.com/weather-history/brazil}{https://www.visualcrossing.com/weather-history/brazil}.}
  \label{fig:climate_loss}
\end{figure}

 \onecolumngrid

 \twocolumngrid

  \onecolumngrid

\begin{figure}[H]
\begin{center}
  \includegraphics[scale = 0.54]{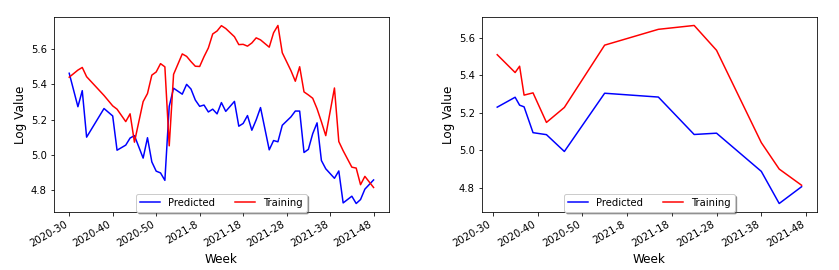}
  \end{center}
  \caption{Actual data and the predicted data  for Brazil when both holiday and climate factors are considered on the training set (Left) and the test set (Right).}
  \label{fig:climate_train}
\end{figure}

 \twocolumngrid

 Once  the parameters for temperature and humidity are added, the loss curve of \autoref{fig:climate_loss}   decreases more gradually. In contrast, the loss curve from   \autoref{fig:holiday_loss} decreased very rapidly.
The actual training data and the values predicted by the Keras model are then plotted in   \autoref{fig:climate_train}. In particular, in   \autoref{fig:climate_train} (Left) the prediction has the same general peaks and valleys as the actual data. However,  whilst the prediction seems to have roughly the same amplitude as the data,   the prediction is shifted down approximately 0.2 on the logarithmic scale. In   \autoref{fig:holiday_test} (Left),  the prediction had roughly the same mean value and a smaller amplitude.
 The model predictions with the actual test data are plotted in   \autoref{fig:climate_train} (Right): compared with   \autoref{fig:holiday_test} (right), the predictions in better   seem to correlate with the trend of the actual data. This graph resembles \autoref{fig:climate_train} (Left) in that the blue line has the same approximate trend as the red line, but the predictions  has a smaller numerical values.

\section{The efficacy of the Correlation Model} \label{Repeating the Model} 

We shall dedicate this section to the study of how well our Correlation Model can serve to understand data from other South American countries. For this,  the model shall be re-trained for datasets from Peru and Colombia. Then, we shall perform a reverse study, flipping the parameters in order to predict the number of Dengue cases on countries that do not have Dengue data readily available such as  Cambodia and Kenya, countries from Southeast Asia and Africa that have significant Dengue.

\subsection{  South American Nations} \label{Repeating the Model on Other South American Nations}

Now that we have the model tested on Brazil, we can use it on other nations as well to see if the model predictions match up with the actual data. The procedure shall be repeated for Peru and Colombia, which are both countries where both Dengue and COVID-19 are very prevalent and frequent. 

\subsubsection{The Correlation Model  on Per\'u} \label{Peru}

According to the \href{https://www3.paho.org/data/index.php/en/mnu-topics/indicadores-dengue-en/dengue-nacional-en/252-dengue-pais-ano-en.html}{Pan-American Health Organization Dataset on Dengue}, Peru had 49,274 cases of Dengue Fever in 2021 and has had 20,491 cases of Dengue during January-June 2022, setting Peru as the second country with the highest Dengue cases in South America, just after Brazil. Therefore, it would be meaningful to see what correlates with the number of Dengue or COVID-19 cases in this country. In order to study the influence of holidays within the dataset from Per\'u, we replicated our Brazilian study but with the new set of national holidays, a list of which has been included in Appendix \ref{appendix_1}. 

Because both Peru and Brazil are South American countries, they celebrate very similar national holidays allowing us to do further comparisons of the datasets.  Furthermore,  for our climate variables we consider the data from \href{https://www.visualcrossing.com/weather-history/Lima/metric/2020-01-01/2022-04-01}{Visual Crossing} to obtain data on humidity, temperature, and precipitation. Since Lima is the capital and largest city, with around one-third of the nations' population, we have chosen it  for our climate data, allowing us to derive the loss curve and prediction curve for Peru shown in \autoref{fig:Peru_loss} and \autoref{fig:Peru_train} below. 

\begin{figure}[H]
\begin{center}
  \includegraphics[scale = 0.55]{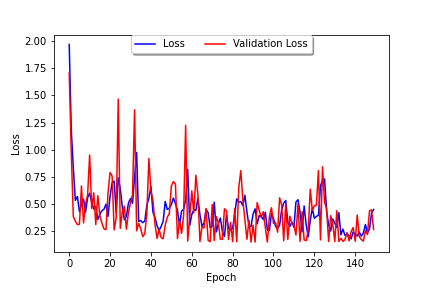}
  \end{center}
  \caption{Loss curve of the neural network model for Per\'u}
  \label{fig:Peru_loss}
\end{figure}

 \onecolumngrid

\begin{figure}[H]
\begin{center}
  \includegraphics[scale = 0.5]{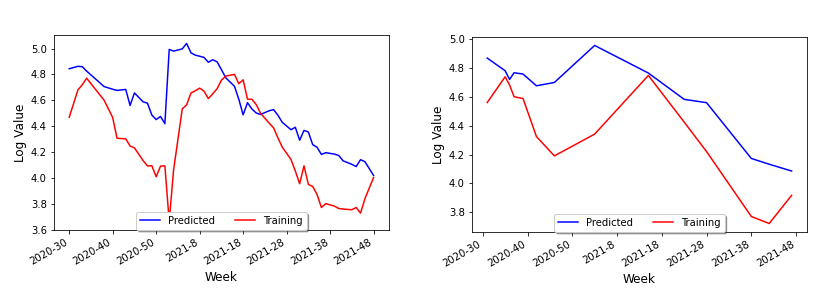}
  \end{center}
  \caption{A graph of the actual data and the predicted data for Per\'u, for when both holiday and climate factors are considered on the training set (Left) and the test set (Right).}
  \label{fig:Peru_train}
\end{figure}

 \twocolumngrid

\subsubsection{The Correlation Model  on Colombia} \label{Colombia}

We shall finally replicate our method for Colombia's dataset.In this case, however, because of incomplete data for Dengue from approximately week 30 to week 60, we shall only focus on the period between week 60 to week 100, and consider the national holidays, a list of which has been included in Appendix \ref{appendix_1}. 
\begin{figure}[H]
\begin{center}
  \includegraphics[scale = 0.45]{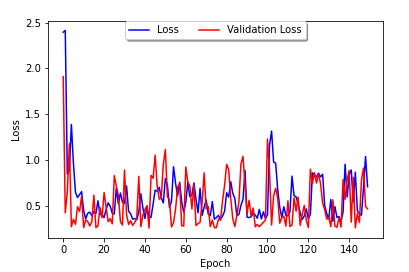}
  \end{center}
  \caption{Loss curve   for Colombia.}
  \label{fig:Colombia_loss}
\end{figure}

Following a similar approach as in the previous sections, we derive the loss curve for Colombia's Correlation Model as shown in \autoref{fig:Colombia_loss}. 
In order to add the variables corresponding to climate factors, we consider  the data from \href{https://www.visualcrossing.com/weather-history/Bogota,Colombia/metric/2020-01-01/2022-04-10}{Visual Crossing} to obtain data on humidity and temperature across the nation. Since Bogot\'a is the both Colombia's capital and largest city, we chose it to be a representative for our climate data in order to do a parallel study to the one done for Lima in the previous section. Then, by using our Correlation Model both on the training set and on the test set, we are able to evaluate its accuracy in predicting infection cases across Colombia as shown in  \autoref{fig:Peru_train}. 

It should be noted that for Brazil, Peru and Colombia, the abrupt drop in cases between weeks   52 y 104 could be due to the low reporting rate associated with the new year's celebrations, which has been seen when studying other diseases \cite{under}. Within our setting, we attempted to mitigate such effect by taking weekly averages when performing our studies.   Finally, it is also important to note that Dengue data  has sometimes been thought to have been underreported, and in particular to have been undermined by the beginning of the COVID-19 pandemic \cite{under2},  and this is why it is sometimes useful to leave the first weeks of the data unused.
\onecolumngrid

\begin{figure}[H]
\begin{center}
  \includegraphics[scale = 0.55]{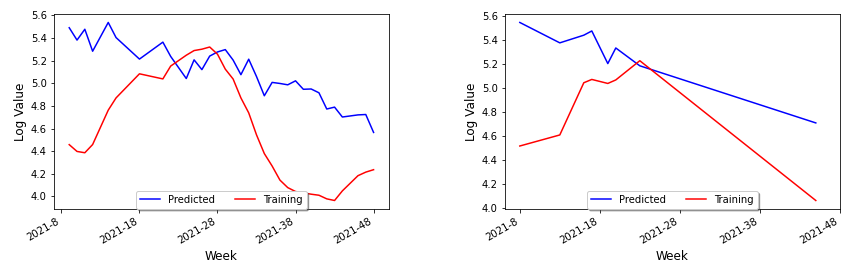}
  \end{center}
  \caption{A graph of the actual data and the predicted data for Colombia, for when both holiday and climate factors are considered on the training set (Left) and the test set (Right).}
  \label{fig:Peru_train}
\end{figure}

 \twocolumngrid


\section{Using the Model to Predict infections   in Africa or Southeast Asia} \label{Using the Model to Predict for Countries in Africa or Southeast Asia}

Having developed our Correlation Model and trained it with the datasets from   several South American countries, we can use the model   to predict the number of Dengue cases for countries in Africa or Southeast Asia, especially since those regions also have a high number of Dengue cases. In particular, whilst not much data can be found on recent Dengue cases for those countries, it is known that the regions of Southeastern Asia have a significant amount of cases of Dengue fever, which creates a heavy economic burden on those countries \cite{SE_Asia}. Thus, by  \emph{reversing} the Correlation Model to use COVID-19 as a parameter to predict Dengue cases over time, one can obtain very useful information for future policy making.

\subsubsection{Predicting Dengue cases in Cambodia} \label{Cambodia}

 We shall first consider Dengue in Cambodia, a Southeastern Asia country. In a recent update as of April of 2022, there have been a total of 457 cases, including one death, which remains much lower than the normal range of cases from 2015 to 2021 \cite{wp_Dengue_Update}. In our model, we shall use South American countries for the training set, and then Cambodia for the test set.

It should be noted that many of the holidays celebrated in Cambodia are quite different from those celebrated in South American countries, and thus we have included the new corresponding list  in Appendix \ref{appendix_1}.
In addition, we shall also include a climate variable built through the data for Cambodia  found in \href{https://www.visualcrossing.com/weather-history/Phnom\%20Penh/metric/2020-01-01/2022-04-16}{Visual Crossing}. Following the same style as before, we shall take Phnom Penh,   the capital and largest city in Cambodia as our city of study.
Finally, it should be noted that instead of using  the log of the number of cases, we shall perform our study by considering the parameter
\[\log\left(\frac{\text{Number of COVID-19 or Dengue cases}}{\text{Country Population in millions}}\right)\]

\begin{figure}[H]
\begin{center}
  \includegraphics[scale = 0.58]{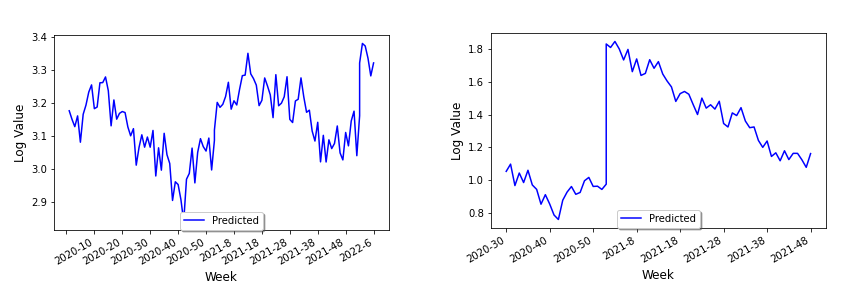}
  \end{center}
  \caption{Predicting Cambodia's Dengue cases using Peru as the training set (Left), and using both Peru and Brazil's data as training set (Right).}
  \label{fig:Cambodia_Peru}
\end{figure}

This kind normalization,   dividing by the nation's population,  gives a better idea of the prevalence of that disease.
By taking Peru as the country for the training set, one obtains the following   prediction of the number of Dengue cases in Cambodia with respect to time shown in \autoref{fig:Cambodia_Peru} (Left). Moreover, by adding  Brazil into the training set,  one can confirm  similar trends with the previous figure, and around the same times of year, as shown in \autoref{fig:Cambodia_Peru} (Right).

\begin{figure}[H]
\begin{center}
  \includegraphics[scale = 0.42]{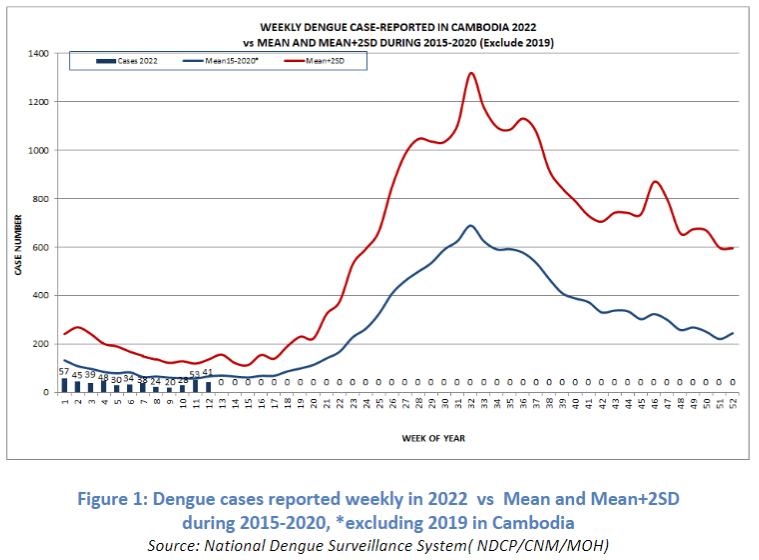}
  \end{center}
  \caption{World Health Organization's Graph of Dengue in Cambodia from 2015-2020}
  \label{fig:Cambodia_update}
\end{figure}

It becomes interesting to compare the above analysis with the average trend of Dengue fever in Cambodia from 2015-2020, from WHO's April 2022 Update of Dengue in the Western Pacific \cite{wp_Dengue_Update}. In particular, one should note the similarities in trend, even though the peaks seen in Figure \autoref{fig:Cambodia_Peru} (Right) and \autoref{fig:Cambodia_update}, which  correspond to different times of the year.\\

 In what follows we shall perform a similar study using Kenya's data set, and include an analysis of such study in the last section of the paper. 

\onecolumngrid

 \twocolumngrid

 \onecolumngrid

 \begin{figure*}
 \begin{center}
  \includegraphics[scale = 0.5]{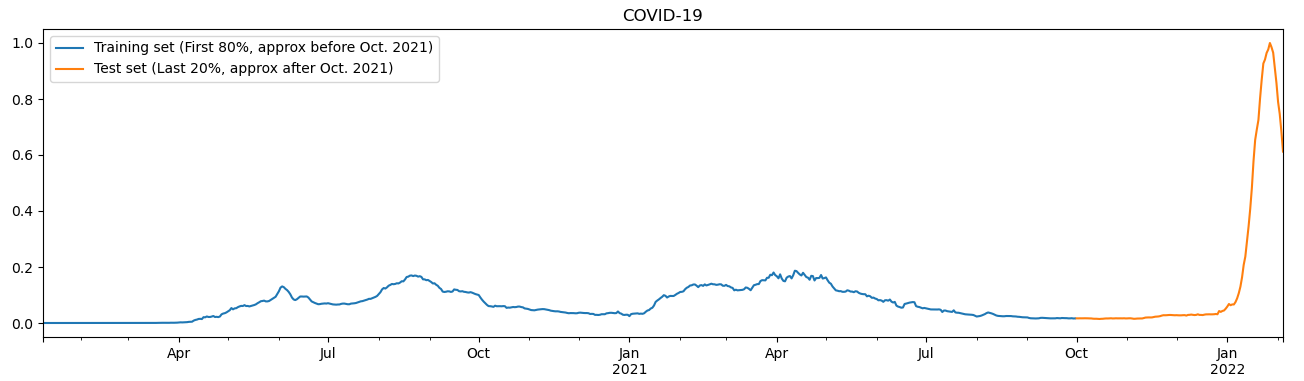}
  \end{center}
  \caption{A scaled look at COVID-19 numbers over the past two years.}
  \label{fig:Scaled_COVID_rawData}
\end{figure*}
 \twocolumngrid

\subsubsection{Predicting Dengue cases in Kenya} \label{Kenya}

\begin{figure}[h]
\begin{center}
  \includegraphics[scale = 0.63]{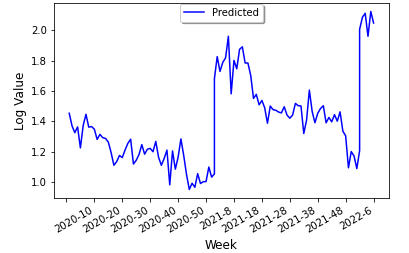}
  \end{center}
  \caption{Predicting Kenya's Dengue cases using Peru and Brazil as the training set.}
  \label{fig:Kenya_predictions}
\end{figure}

In what follows we shall continue showing the utility of our model by considering the data sets for Kenya, one of the countries in Eastern Africa with  a significant amount of Dengue cases. By applying our Correlation Model trained with South American countries on Kenya, one can   predict infection peaks at roughly epidemiological weeks 60 and 110 and valleys at epidemiological weeks 50 and 100 as shown in \autoref{fig:Kenya_predictions}.

Having seen how many external variables, such as celebration of holidays or temperature or humidity, affect the prevalence of COVID-19 or Dengue, in what follows, we shall expand on our Correlation Model and build a Recurrent Neural Network and perform a time series analysis (see Section \ref{LSTM_intro} for background on LSTM models). 

In what follows, it is useful to have in mind  the number of COVID-19 infections from the beginning of 2020 up until the beginning of 2022 as given in \autoref{fig:Scaled_COVID_rawData}, with data used from 0 -- 1.

\pagebreak
\section{Recurrent Neural Network and Time Series Analysis} \label{LSTM}

We shall dedicate this section to expand our Correlation Model to build upon a Recurrent Neural Network. To train this new model, we shall use  Per\'u's dataset, since it has the most complete data for Dengue and thus provides an ideal starting point. Moreover, it should be noted that    the number of epochs that one chooses for the model has a large  impact on the model: the smaller the number of epochs, the more the model is likely to under-fit; Equivalently, the larger the number of epochs, the more the model tends to over-fit. In what follows we shall will assign the first 80\% of the data, roughly until October 2021, for the training set and the remaining 20\% for the test set, and use MinMaxScaler, so that the calculations are done with all the data   from 0 to 1.

Through the data of  \autoref{fig:Scaled_COVID_rawData}, we can  inverse of our  original rescaled model first, and then consider the base 10 logarithm  for all of the values, to obtain a prediction of  COVID-19 infections using a time series analysis for Peru as shown in \autoref{fig:Peru_COVID_log_firstHalf} below. 

\begin{figure}[h!]
  \includegraphics[scale = 0.52]{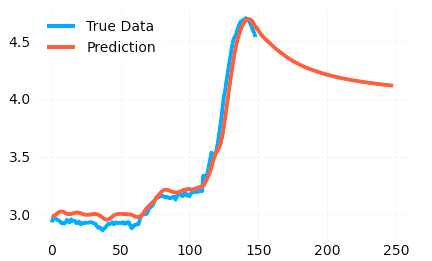}
  \caption{Predicting COVID-19 infections using time series.}
  \label{fig:Peru_COVID_log_firstHalf}
\end{figure}
It should be noted that   the horizontal scale is measured in days, and thus from our model one expects the number of COVID-19 cases to decrease gradually after the spike in January of 2022, which is reasonable considering that Omicron caused a large increase in the number of cases.
 
\begin{figure*}
  \includegraphics[scale = 0.4]{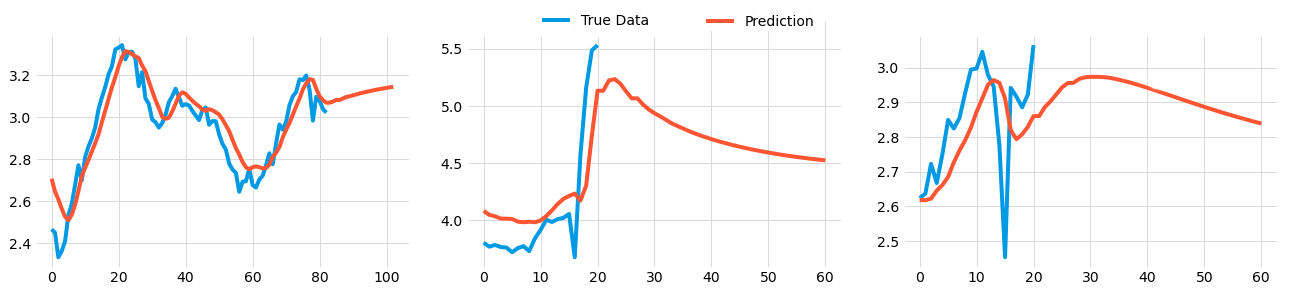}
  \caption{Predictions of infections using a time series analysis and LSTM model for Dengue (Left), COVID-19 (middle), and for predicting Dengue infections using an LSTM model that incorporates both diseases (Right).}
  \label{fig:Peru_Dengue_log_LSTM_pred}
\end{figure*}

A similar study can then be done for the prediction of Dengue and COVID using LSTM, which we shall illustrate in  Figure \ref{fig:Peru_Dengue_log_LSTM_pred} above.  In the case of Dengue, one can see in \ref{fig:Peru_Dengue_log_LSTM_pred} (Left) that  with enough epochs the prediction fits the true data quite well.

  The case of Dengue infections has been considered in \autoref{fig:Peru_Dengue_log_LSTM_pred} (Middle), where the blue spike downwards is actually probably due to a missing value of sorts and may not be an actual zero. Considering this, the model gets most of the general trends pretty accurately. After the beginning of 2022, this model predicts that the number of Dengue infections will initially rise and then fall, which we can compare with   \autoref{fig:Peru_Dengue_log_LSTM_pred} (Left), in which we have incorporated both epidemics. Essentially, there are two models here; one using past values for both diseases to predict COVID-19, and the other using past observed values for both to predict Dengue. These are updated simultaneously while the program is running. Again, we have used MinMaxScaler to scale the numbers from 0 to 1, then inverse scaled them back and took the logarithm of all value to make the details more clear.

\section{Conclusion and applications} \label{Additional Ideas}

In this paper we have built a neural network Correlation Model as well as a recurrent Correlation model for a neural network to take advantage of the correlation between Dengue and COVID-19 to deduce predictions on future infections of both diseases. Such correlations drawn should also present interesting lines of research for other viruses which present correlations. Moreover, there are many directions in which our models could further be expended. Since it is outside the scope of this paper to perform those extensions here, we shall instead give a brief summary of the directions we believe would be very fruitful to explore in upcoming work.

 In particular,  we shall consider  Latent Heat Flux and whether it could be a better correlator than other climate factors, such as humidity, temperature, or precipitation. Finally, we shall also propose  a notion of peak indices which could be the ideal setting when finding correlations between viral epidemics. Finally, we shall conclude this note by summarizing our methods and highlighting our findings. 
 \hfill
 \pagebreak
 
 \subsection{Latent Heat Flux} \label{Latent Heat Flux}
  From our model building analysis, it would seem that  the time series of transmissions of COVID-19 and Dengue could be better correlated to the latent heat flux (LHF), which is the amount of energy moving between the air and the land due to evaporation or condensation. In particular, the research of heat and mass transfers has been identified as a possible area that could lead to progress in reducing COVID-19 \cite{thermo_correlation}. A sufficient dataset on latent heat flux could not be found because the datasets   that are available online stopped at the early-2010's and were not daily or weekly, but there are many environmental factors correlated with LHF or just COVID-19 and Dengue which could be taken into consideration Instead. Some of the ones we believe could be best suited for finding correlations are the following:
 \begin{itemize}
 \item Sea Surface Temperature \cite{SST} -- unfortunately, no data on this either;
 \item Temperature change;

 \item Humidity -- the higher the humidity, the higher the latent heat flux tends to be;
 \item Wind speed -- the greater the wind speed, the lower the latent heat flux usually is; 
 \item Atmospheric pollution impacts \cite{thermo_correlation} -- for this, use the Visibility data from the Visual Crossing dataset.
 
  \end{itemize}
 
\begin{figure*}
    \centering
\includegraphics[width=\textwidth]{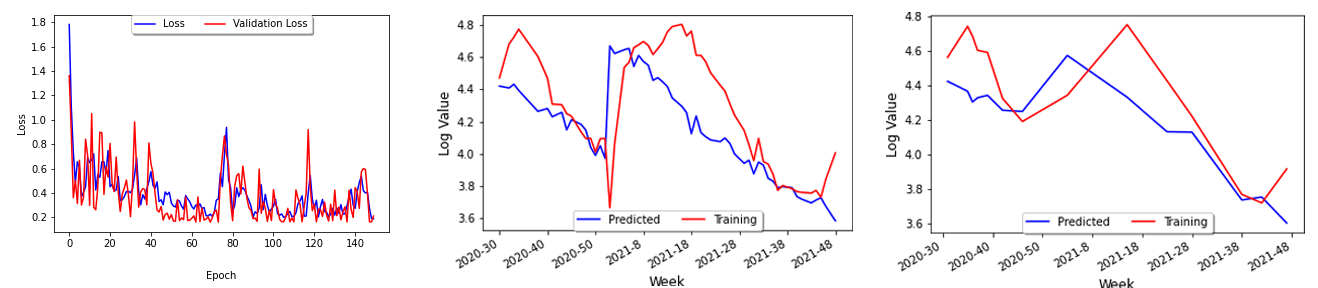}
    \caption{(Left) Loss curve (Center) Training predictions (Right) Test predictions}
    \label{fig:Peru_B}
\end{figure*}

From the above, one can add the new parameters of 
\begin{itemize}
\item maximum temperature, 
\item minimum temperature, 
\item wind speed, and
\item  visibility
\end{itemize} 
for the study of Per\'u's dataset (since that was the country that had the most complete data) leading to the graph of the loss curve and the prediction models shown in \autoref{fig:Peru_B} shown in the following page.

 \onecolumngrid

\newpage

  \twocolumngrid

\subsection{Peak Index Predictions} \label{Peak Index Predictions}
It should be very useful to understand correlations between viral epidemics through some sort of   \emph{peak index}, which can be defined in a general sense as the week index of where the largest number of COVID-19 or Dengue cases occur. In particular, the following two perspectives could be incorporated to our Correlation Model to expand on our work: 

\begin{itemize}
\item{\bf  A relative index}. We define an epidemiological week to be a peak if it has a greater number of cases than both the week before and after it. We define a function $\rho (t)$ to be the function that represents whether EW $t$ is a peak. Thus,
\begin{eqnarray}
   \rho (t) &=& 
   1 ~{\rm if} ~ Nc[t] \ge \alpha Nc[t-1] ~{\rm and} Nc[t] \ge \alpha Nc[t+1]\nonumber\\
     &=&0 ~{\rm otherwise} ~
\nonumber
\end{eqnarray}

\item {\bf An absolute index. } We define a function $\rho (t)$ to be the function that represents whether EW $t$ is a peak. Formally, this is given by
\begin{eqnarray}\rho (t) &=&    1~{\rm  if}~ Nc \geq \alpha  \cdot  P\nonumber\\
  &=& 0 ~{\rm otherwise},  \nonumber\end{eqnarray}
where $Nc$ represents the number of new infections for a particular disease and $P$ represents the total population for a particular country, and  $\alpha$ is a coefficient that differs among various diseases. For example, for COVID-19, this could be if the new cases is at least $\frac{1}{10,000}$ of the country's population. For Dengue, this could be if the number of new cases is at least  $\frac{1}{1,000,000}$ of the country's population (since there are currently far less cases for Dengue than for COVID-19)
\end{itemize}

 \subsection{Final remarks}\label{conclusion}
 By considering datasets from different South American countries on COVID-19 and Dengue cases, we have been able to build a Correlation Model which allows us to study COVID-19 and Dengue cases in areas of the world where data is not fully available. In particular our study,  we have been able to observe the following. 
 \begin{itemize}
 \item When considering the importance of holidays as a parameter, the prediction curve has approximately the same mean as the actual data and yet has a smaller amplitude, and it seems as if the prediction essentially diminished the variance of the initial data as seen in \autoref{fig:holiday_test} (Left). 
 \item When  parameters for temperature and humidity are added, the loss curve seems to decrease more gradually as shown in \autoref{fig:climate_loss}. In contrast, the loss curve from Figure \ref{fig:holiday_loss} decreased very rapidly.
 \item By only adding in climate factors into the neural network results in a gradually descending loss curve, but with very big amplitude as shown in \autoref{fig:holiday_loss}. Combining this with holiday factors seems to take care of this issue.
 \item When considering combined holiday and climate factors, the prediction seems to have roughly the same amplitude as the data, but the prediction is shifted down approximately 0.2 on the logarithmic scale as shown in \autoref{fig:holiday_test} (Left). However, interestingly  the prediction had roughly the same mean value and a smaller amplitude.
 \item The sequence length is also a highly important factor that may change the prediction: indeed,  when changing the 14-day sequence length to 35-days, the prediction for COVID-19 is to decrease (like before) and then rise slowly.
 \end{itemize}

 Finally, it would be interesting to see the effects of the COVID-19 lockdown on Dengue \cite{reducer_effect}. Many say that because the lockdown occurred so close to the peak of the mosquito season, the transmission of Dengue increased \cite{Asia_Dengue}, as the lockdown affected \emph{Aedes} mosquito surveillance and vector control procedures \cite{DANIELREEGAN2020e05181}. In this setting, then mosquito larvae populations would have increased,  and thus the number of people infected with Dengue would increase because they lived in close proximity with mosquitos.   It should be noted that  the number of Dengue cases could have been severely underestimated, simply due to the fact that health resources were being focused on COVID-19 instead or that people who were infected with Dengue were unwilling to get a formal diagnosis due to the lockdown \cite{epidemiologia3010009}.

 \noindent {\bf Acknowledgments.}
  Grace Wang and Laura P. Schaposnik would like to express their deep gratitude to the MIT PRIMES Program for this wonderful opportunity to conduct research, and to Fidel A. Schaposnik for introducing them to the work of Paula Bergero which led to this collaboration.   \bigskip
 
 \noindent {\bf  Contributions of authors.} All authors contributed equally to the planning of the paper. Much of the work was done by Grace Wang under the supervision of Laura P. Schaposnik through PRIMES-MIT program 2022. 
 \newpage
 
 \noindent {\bf Funding.} 
 PB  is member of the Scientific Career of Consejo Nacional de Investigaciones Cientí\'iicas y T\'ecnicas of Argentina (CONICET). The work of LPS is partially supported by  a Simons Fellowship, NSF CAREER Award DMS 1749013 and NSF FRG Award 2152107.
 
 \bigskip
 
 \noindent {\bf Affiliations.}\\
 (a)~ Universidad Nacional de La Plata $\&$ CCT, Argentina. \\
 (b)~ University of Illinois, Chicago, USA. \\
 (c)~ PRIMES-MIT, USA. 
 
 \bigskip
   
 \bibliography{COVID2022}{}
 \bibliographystyle{fredrickson}
 

%
%
%
%



 \newpage
 \appendix
 
 \section{Holidays Used} \label{appendix_1}
 \begin{small}

Brazil Holidays:

\begin{itemize}
\item New Year's Day (January 1)
\item Shrove Tuesday (March 1)
\item Good Friday and Easter (around April 12, 2020 and April 4, 2020)
\item Tiradentes Day (April 21)
\item Labour Day (May 1)
\item Feast of Corpus Christi (June 16)
\item Independence Day (September 7)
\item Our Lady of Aparecida's day (October 12)
\item All Souls Day (November 2)
\item Republic Day (November 15)
\item Christmas Day (December 25)
\item Brazilian Carnival -- February 21 - 26 (in 2020) and February 12 - 17 (in 2021)
\end{itemize}

Peru Holidays:

\begin{itemize}
\item New Year's Day (January 1)
\item Good Friday and Easter Sunday (April 2 and April 4 in 2021)
\item Labour Day (May 1)
\item Saint Peter and Saint Paul (June 29)
\item Independence Day (July 28-29)
\item Santa Rosa de Lima (August 30)
\item Battle of Angamos (October 8)
\item All Saints' Day (November 1)
\item Immaculate Conception Day (December 8)
\item Christmas Day (December 25)
\end{itemize}

Colombia Holidays:

\begin{itemize}
\item New Year's Day (January 1)
\item Epiphany (January 10)
\item St. Joseph's Day (March 21)
\item Maundy Thursday and Good Friday (April 14 and April 15 in 2021)
\item Labour Day (May 1)
\item Ascension Day Holiday (May 30)
\item Corpus Christi Holiday (June 20)
\item Sacred Heart (June 27)
\item Saints Peter and Paul's Day (July 4)
\item Declaration of Independence (July 20)
\item Battle of Boyaca (August 7)
\item Assumption Day (August 15)
\item Day of the Races (October 17)
\item All Saints' Day Holiday (November 7)
\item Independence of Cartagena (November 14)
\item Immaculate Conception Day (December 8)
\item Christmas Day (December 25)
\end{itemize}

Cambodia Holidays:

\begin{itemize}
\item New Year's Day (January 1)
\item Victory over Genocide Day (January 7)
\item International Women's Day (March 8)
\item Khmer New Year (April 14-16 in 2021 and August 17-21 in 2020, as a replacement holiday)
\item Visak Bochea Day (April 26)
\item Royal Plowing Ceremony (April 30)
\item International Labour Day (May 1)
\item King's Birthday (May 14)
\item King's Mother's Birthday (June 18)
\item Constitutional Day (September 24)
\item Ancestors' Day (October 5-7)
\item Commemoration Day of King's Father (October 15)
\item King's Coronation Day (October 29)
\item Independence Day (November 9)
\item Water Festival Ceremony (November 18-20)
\end{itemize}

Kenya Holidays:

\begin{itemize}
\item New Year's Day (January 1)
\item Good Friday and Easter Monday (April 15 and April 18, respectively)
\item President Kibaki State Funeral (April 19)
\item Labour Day (May 2)
\item Idd-ul-Fitr (May 3)
\item Madaraka Day (June 1)
\item Idd-ul-Azha (July 10)
\item Utamaduni Day (October 10)
\item Mashujaa Day (October 20)
\item Jamhuri Day (December 12)
\item Christmas Day and Boxing Day (December 25 and December 26, respectively)
\end{itemize}
 
\end{small}
\end{document}